\documentclass{aastex62}

\graphicspath{{./}{figures/}}

\received{May 3, 2018}
\submitjournal{ApJ}

\shorttitle{\emph{Gaia} L dwarf discovery}
\shortauthors{Faherty et al.}

\begin{document}

\title{A Late-type L Dwarf at 11 pc Hiding in the Galactic Plane Characterized Using \emph{Gaia}~DR2}

\correspondingauthor{J. Faherty}
\email{jfaherty@amnh.org}

\author[0000-0001-6251-0573]{Jacqueline K. Faherty}
\affil{Department of Astrophysics, American Museum of Natural History, Central Park West at 79th Street, NY 10024, USA }

\author[0000-0002-2592-9612]{Jonathan Gagn\'e}
\affiliation{Carnegie Institution of Washington DTM, 5241 Broad Branch Road NW, Washington, DC~20015, USA}

\author[0000-0002-6523-9536]{Adam J. Burgasser}
\affiliation{Center for Astrophysics and Space Sciences, University of California, San Diego, San Diego, CASS, M/C 0424 9500 Gilman Drive, La Jolla, CA 92093-0424, USA}
\affiliation{NASA Sagan Fellow}

\author[0000-0003-2008-1488]{Eric E. Mamajek}
\affiliation{Jet Propulsion Laboratory, California Institute of Technology, M/S
  321-100, 4800 Oak Grove Drive, Pasadena, CA 91109, USA}
\affiliation{Department of Physics \& Astronomy, University of
  Rochester, Rochester, NY 14627, USA}
  
\author[0000-0003-4636-6676]{Eileen C. Gonzales}
\affiliation{Department of Astrophysics, American Museum of Natural History, New York, NY 10024, USA}
\affiliation{The Graduate Center, City University of New York, New York, NY 10016, USA}
\affiliation{Department of Physics and Astronomy, Hunter College, City University of New York, New York, NY 10065, USA}

\author[0000-0001-8170-7072]{Daniella C. Bardalez Gagliuffi}
\affiliation{Department of Astrophysics, American Museum of Natural History, New York, NY 10024, USA}

\author[0000-0001-7519-1700]{Federico Marocco}
\altaffiliation{NASA Postdoctoral Program Fellow}
\affiliation{Jet Propulsion Laboratory, California Institute of Technology, M/S 169-237, 4800 Oak Grove Drive, Pasadena, CA 91109, USA}

 \begin{abstract}
We report on the characterization of a nearby ($d$ = 11.2\,pc) ultracool L dwarf (\emph{WISE} J192512.78+070038.8; hereafter W1925) identified as a faint ($G$ = 20.0) high proper motion (0$\farcs$22 yr$^{-1}$) object in the \emph{Gaia} Data Releases 1 and 2. A Palomar/TripleSpec near-infrared spectrum of W1925 confirms a previously estimated photometric L7 spectral type by \citet{Scholz18}, and its infrared colors and absolute magnitudes are consistent with a single object of this type. We constructed a spectral energy distribution using the \emph{Gaia} parallax, literature photometry and near infrared spectrum and find a luminosity log($L_{\rm bol}$/$L_{\odot}$)\,=\,-4.443$\pm$0.008.  Applying evolutionary models, we infer that W1925 is likely a 53\,$\pm$\,18\,$M_{\rm Jup}$ brown dwarf with $T_{\rm eff}$\,=1404\,$\pm$\,71\,K and $\log{g} $\,=5.1$\pm$0.4\,dex (cgs). While W1925 was detected in both the 2MASS and \emph{WISE} infrared sky surveys, it was not detected in photographic plate sky surveys.  Its combination of extreme optical-infrared colors, high proper motion, and location near the crowded Galactic plane ($b$ = -4$\fdg$2) likely contributed to it having evaded detection in pre-\emph{Gaia} surveys. 
\end{abstract}


\section{Introduction} \label{sec:intro}

Despite decades of dedicated searches, the nearby stellar sample ($d \lesssim$ 25\,pc) remains incomplete, particularly for intrinsically faint low-mass stars and brown dwarfs \citep{Henry18}. Some of the nearest systems to the Sun, such as the binary brown dwarf \emph{WISE}~J104915.57$-$531906.1AB at 2.0\,pc (3rd closest system to the Sun; hereafter \emph{WISE}~J1049$-$5319; \citealt{Luhman13}) and the Y dwarf \emph{WISEA}~J085510.74$-$071442.5 at 2.2\,pc (4th closest system to the Sun; \citealt{Luhman14}), were only recently uncovered through multi-epoch infrared imaging surveys (see \citealt{Luhman14b}).  While these surveys, including the Two Micron All Sky survey (2MASS; \citealt{Skrutskie06}), the Sloan Digital Sky Survey (SDSS; \citealt{York00}), the UKIRT Infrared Deep Sky Survey (UKIDSS; \citealt{Lawrence07}), and the Wide-field Infrared Survey Explorer (\emph{WISE}; \citealt{Wright10}) have enabled new discoveries in the Solar neighborhood, few have been made in or near the Galactic plane (e.g. \citealt{Burgasser02}, \citealt{Kirkpatrick99},  \citealt{Kirkpatrick16}, \citealt{Kirkpatrick14}, \citealt{Burningham13}, \citealt{Lucas10}, \citealt{Gizis11}, \citealt{Artigau10}, \citealt{Kuchner17}).  Most all-sky searches for nearby, low-temperature dwarfs avoid this region of the sky due to high rates of contamination from heavily reddened background stars with similar colors and astrometric confusion leading to spurious proper motion measurements.  

The nearby stellar sample serves as a benchmark for testing fundamental laws that apply across the Milky Way and beyond. One can extrapolate from the numbers, distributions, and diversity among stars and brown dwarfs in the 20\,pc sample to understand overall Galactic demographics  (e.g. \citealt{Subasavage17}, \citealt{Clements17}, \citealt{Jao17}).  Completing the nearby census remains an important goal in stellar population studies.

In this Letter, we report the independent discovery and characterization of a nearby (11.20$\pm$0.08~pc), low-temperature L-type brown dwarf \emph{WISE}~J192512.78+070038.8 (hereafter W1925). In Section~\ref{sec:ident} we describe the identification of this source in the \emph{Gaia} Data Release 2 (\emph{Gaia}~DR2 hereafter; \citealt{2016A&A...595A...1G})\footnote{\url{https://gea.esac.esa.int/}} and its confirmation as a red, high proper source from multi-epoch survey data. In Section~3 we summarize its existing astrometric and photometric observations and report new infrared spectroscopic observations. In Section~4 we analyze these observations, inferring the spectral type, location on color and absolute magnitude diagrams, kinematics, and fundamental properties of the source. In Section~5 we place this source in context with the nearby stellar sample and the potential for future discoveries in the Galactic plane.

\section{Identification of W1925}\label{sec:ident}
\emph{Gaia}~DR2 was released on 2018 April 25 (\citealt{GaiaDR2}), and contains 1,692,919,135 point sources, of which 1,331,909,727 have a five parameter astrometric solution (position, parallax, and proper motions) and 7,224,631 have a 5 parameter solution plus a radial velocity measurement. 

\begin{figure*}[!h]  
\begin{center}
\epsscale{1.0}
\plotone{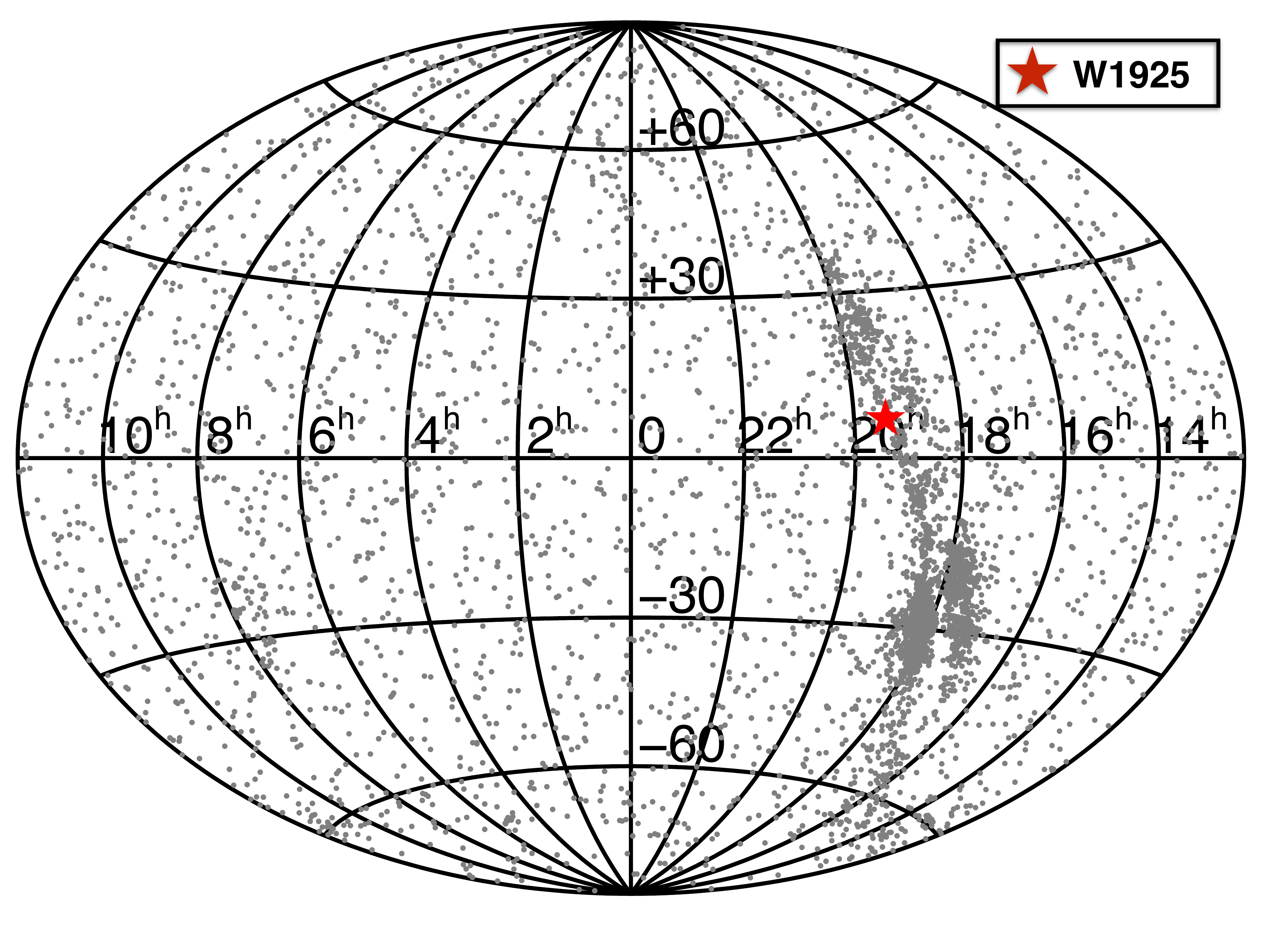}
\end{center}
\caption{An Aitoff projection sky chart of the equatorial positions ($\alpha$, $\delta$; ICRS) of all \emph{Gaia}~DR2 stars reported to have a parallax $>$\,50\,mas ($d$\,$<$\,20\,pc). No astrometric quality cuts have been applied to the data.  The Galactic plane can be clearly distinguished, the bulk of which is contamination.  W1925 was recovered near the plane ($b$\,=\,$-4\fdg$2; red star). \label{fig:projection}} 
\end{figure*}

We promptly began searching the data for the reddest, new sources with robust parallax measurements.  We downloaded the 20\,pc ($\varpi$\,$>$\,50\,mas) sample from the \emph{Gaia} archive, which includes 5400 point sources.  The vast majority of these sources align with the southern Galactic plane as shown in Figure ~\ref{fig:projection}, 
indicating that they are largely false detections.  Nevertheless, as nearly all previous searches for brown dwarfs avoided the Galactic plane due to crowding, we did not immediately reject any source. Instead, we calculated absolute \emph{Gaia} $G$--band magnitudes and examined only sources with $M_{G}$\,$>$\,17.0 (2,859 sources) to focus on the lowest-mass stars and brown dwarfs.  

The \emph{Gaia}~DR2 catalog comes with quality indicators that allow vetting of targets with poor astrometric solutions (e.g. the contaminants in the galactic plane).  For instance,  \texttt{VISIBILITY\_PERIODS\_USED} indicates the number of distinct observation epochs, \texttt{ASTROMETRIC\_EXCESS\_NOISE} is the excess source noise, and the \texttt{ASTROMETRIC\_5DSIGMA\_MAX} parameter is a five-dimensional equivalent
to the semi-major axis of the position error ellipse. We examined all of these quality indicators to ascertain if they might help vet nearby targets in dense regions. In the case of the Galactic plane we find many objects have $<$\,10 visibility periods used.  However, we also find that Proxima Centauri, the closest star to the Sun, has only 8 visibility periods used.  Therefore we decided not to apply rejection criteria based on quality indicators alone.  We list the most relevant \emph{Gaia}~DR2 quality indicators for W1925 in Table ~\ref{tab:param}.

\begin{figure*}[!h] 
\begin{center}
\epsscale{1.0}
\plotone{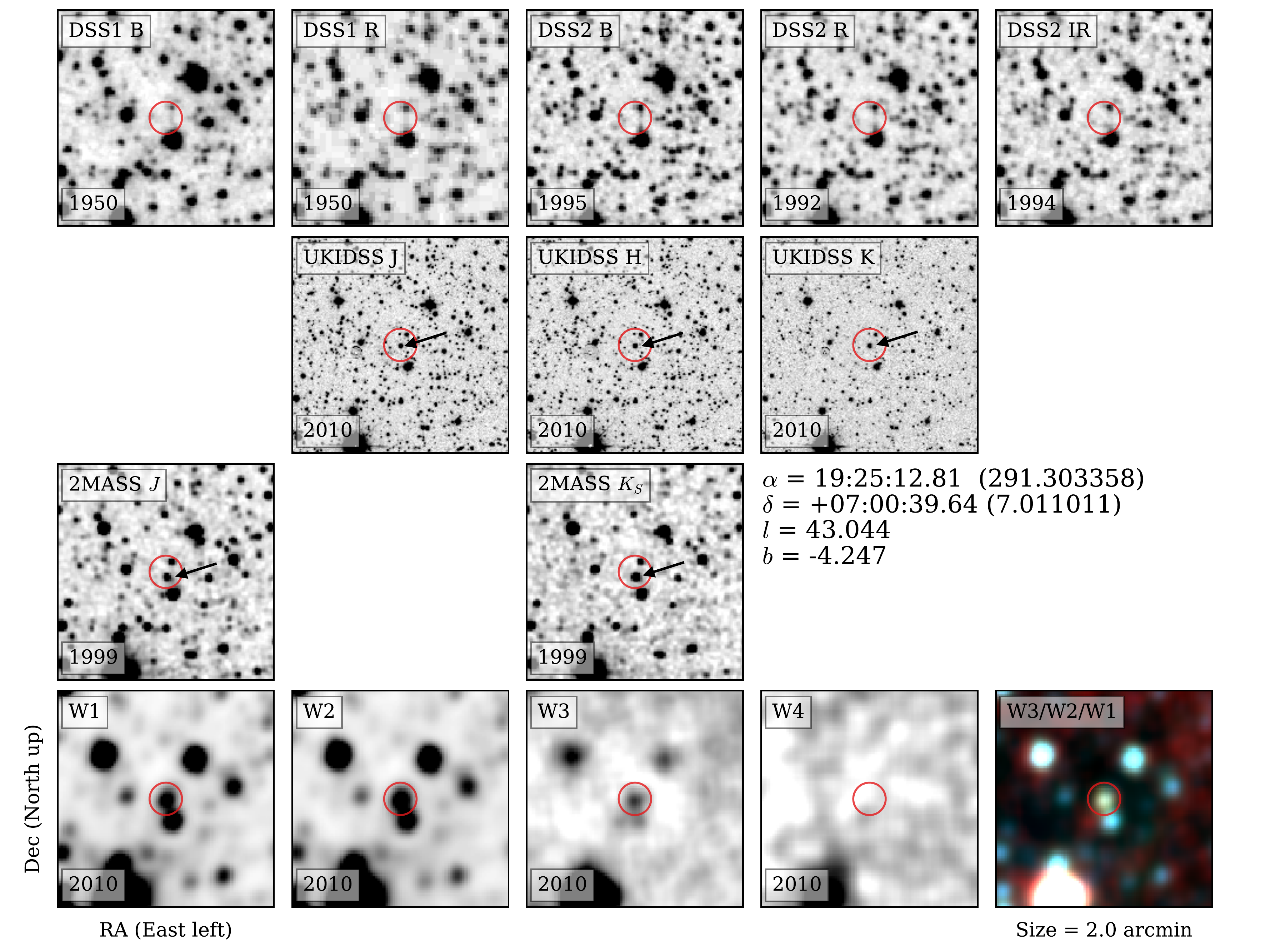}
\end{center}
\caption{A finder chart for W1925 centered on the current \emph{Gaia} reported position (red circle).  We show 2.0 arcmin boxes around DSS, UKIDSS, 2MASS, and \emph{WISE} images and find W1925 clearly detected in the latter three.  We have used an arrow to distinguish the position of W1925 in multiple images.\label{fig:finder}} 
\end{figure*}

Instead, we filtered out contaminants by projecting all \emph{Gaia} epoch 2015 coordinates within 20\,pc backwards to epoch 2000.0, and searched for a genuine detection in 2MASS within 2$\arcsec$ of the new position. Our goal was to observe new 20\,pc candidates with upcoming spectroscopic time, so we limited our search to 331 sources observable in April-May 2018. We visually examined images from the Digital Sky Survey (DSS), Pan-STARRS (where available; \citealt{2002SPIE.4836..154K}), 2MASS and \emph{WISE} using the Aladin tool (see also Figure ~\ref{fig:finder}).  Among these sources, we identified 41 known L and T dwarfs.  

We identified W1925 (source ID \emph{Gaia}~DR2 4295524821431807232) as a faint, red object, with detections in the 2MASS (2MASS~J19251275+0700362), Pan-STARRS (J192512.79+070039.0), \emph{WISE} (WISE~J192512.78+070038.8) and AllWISE (WISEA~J192512.77+070038.6) catalogs based on its proper motion. The source had previously been reported by \citet{Scholz18} as a potential nearby candidate L dwarf in \emph{Gaia} DR1, but without spectroscopic confirmation.  Figure~\ref{fig:finder} shows images of the sky around this source in DSS, UKIDSS, 2MASS and WISE, with the last two revealing a clearly moving source.  The faint absolute magnitude ($M_{G}$\,=\,19.79\,$\pm$\,0.02) and red optical-infrared color ($G-K_s$\,=\,7.10\,$\pm$\,0.04) of W1925 indicates it to be a low-luminosity, low-temperature source. 
   
\section{Observations}

\subsection{Astrometry and Photometry}

Table~\ref{tab:param} lists all published photometry and \emph{Gaia}~DR2 astrometry for W1925. While it is detectable in the red optical bands of Pan-STARRS, there are no entries for this source in GSC 2.2 or USNO-B1.0 catalogs, and it does not appear in any of the DSS photographic plates scanned by SuperCOSMOS (\citealt{2001MNRAS.326.1279H}; see the first three panels of Figure ~\ref{fig:finder}). 

The exceptional optical-infrared color is likely responsible for the absence of this source in early optical images despite its relative brightness in the infrared ($K_s$\,=\,12.94\,$\pm$\,0.04).

\subsection{Spectroscopy}
W1925 was observed with the TripleSpec near-IR spectrograph \citep{2004SPIE.5492.1295W,Herter08} on the Palomar 200$''$ telescope on 2018 April 28 UT.  TripleSpec covers 1.0--2.4\,$\mu$m at a resolution R\,$\sim$\,2600 with a fixed 1\arcsec\ $\times$ 30\arcsec\ slit. Conditions during the observation were clear and dry, with seeing around 1.0\arcsec-1.4\arcsec. We observed W1925 starting at 11:27~UT at an airmass of 1.183, and obtained 8 frames of 180\,s each in an ABBA nodding pattern for a total integration time of 1440\,s. The slit was aligned with the parallactic angle. Data were reduced using a modified version of SpeXtool \citep{2004PASP..116..362C}, following standard procedures for imaging processing, order identification, and spectral extraction; and wavelength calibration was determined using OH airglow lines (RMS scatter = 5\,km\,s$^{-1}$). Telluric absorption correction and flux calibration were determined from observations of the A0~V star HD~183324 ($V = 5.783$) immediately after the W1925 observation, following the procedures of \citet{Vacca03}.

\section{Analysis}
\subsection{Spectra}
\begin{figure*}[!h]    
\begin{center}
\epsscale{1.0}
\plotone{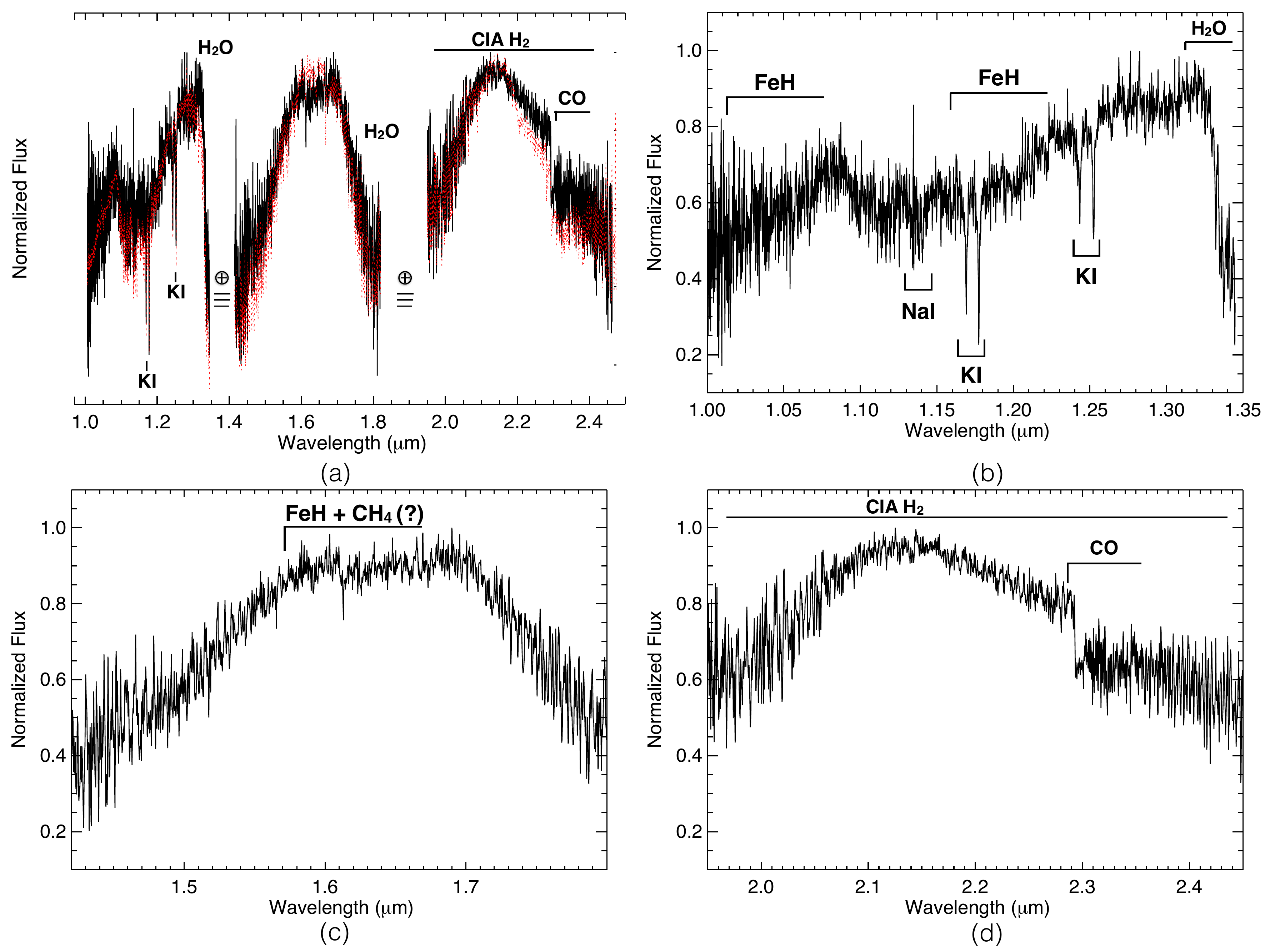}
\end{center}
\caption{The near infrared spectrum of W1925. Panel (a) shows the full spectrum (black line), normalizing each band separately (between telluric bands), and comparing to similar data for \emph{WISE}~J1049$-$5319~A (red line) from \citet{Faherty14}.  Panels (b), (c), and (d) show each of the JHK bands separately, with relevant features labeled.\label{fig:spectra}}
\end{figure*}

Figure ~\ref{fig:spectra} shows the reduced near-infrared spectrum for W1925.  We used the SpeX Prism Library Analysis Toolkit (SPLAT; \citealt{2017arXiv170700062B}) to compare with M, L and T dwarf low resolution near-infrared spectral standards, and found the best overall fit to the L7 standard \emph{2MASSI}~J0103320+193536 \citep{2010ApJS..190..100K}.  In Fig.~\ref{fig:spectra}a, we compare W1925 to similar data from the Folded-port InfraRed Echellette spectrograph (FIRE; \citealt{Simcoe13}) of \emph{WISE}~J1049$-$5319~A (\citealt{Faherty14}), which has a similar L7.5 infrared spectral type (\citealt{Burgasser13}).  There are subtle differences in the spectra, including CH$_{4}$ in H band and the redder shape of the 2.2\,$\mu$m peak; however, overall the spectra are very similar. 

The spectra of L dwarfs are known to have a variety of gravity-sensitive features, including VO and metal hydride molecular bands, alkali lines, and the H-band peak shape \citep{Lucas01,McGovern04,Cruz09,Allers13}. 
 
\citet{Allers13} define a spectral index approach for estimating surface gravity in late M and L dwarfs; however, the method is not well-calibrated for L7 objects. Nevertheless, applying their methodology we find a near infrared spectral type of L7.1 (consistent with our analysis) and gravity indices of \texttt{nn00} implying a \texttt{FLD-G} (field gravity) designation. From visual inspection, we see that W1925 has strong \ion{K}{1} absorption (equivalent widths of 7.1$\pm$0.9~\AA, 8.9$\pm$1.0~\AA\ for the 1.168 , 1.177 $\mu$m doublet and 4.9$\pm$0.6\,\AA, 6.0$\pm$0.7\,\AA\ for the 1.243, 1.254\,$\mu$m doublet) which place W1925 in line with other field L dwarfs \citep{Martin17}. It also has a flattened H-band peak and no evidence of VO absorption at 1.05~$\micron$. These features consistently indicate that W1925 has a surface gravity similar to field brown dwarfs, $\log{g}$ $\approx$ 5.0--5.5\,dex (cgs; see Section~\ref{sec:fundamentals}). 

We used the spectral data to measure the radial velocity of W1925 by cross-correlating regions containing strong features with a high-resolution $T_{\rm eff}$ = 1400\,K, $\log{g}$\,=\,5.0\,dex (cgs) BT-Settl atmosphere model \citep{2012RSPTA.370.2765A}. The comparison regions were 
1.16--1.185\,$\micron$ and 
1.235--1.26\,$\micron$ (\ion{K}{1});
1.32--1.34\,$\micron$,
1.46--1.50\,$\micron$ and
1.75--1.78\,$\micron$ (H$_2$O); and
2.03--2.06\,$\micron$ (\ion{Na}{1} and CO).
The heliocentric radial velocity was determined to be $-$9$\pm$7\,km\,s$^{-1}$, which includes a barycentric velocity correction of +25.4\,km\,s$^{-1}$, and the uncertainty is based on the scatter of measurements from the six comparison regions.

\subsection{Color and Absolute Magnitude Analysis}
\begin{figure*}[!h]  
\begin{center}
\epsscale{1.0}
\plotone{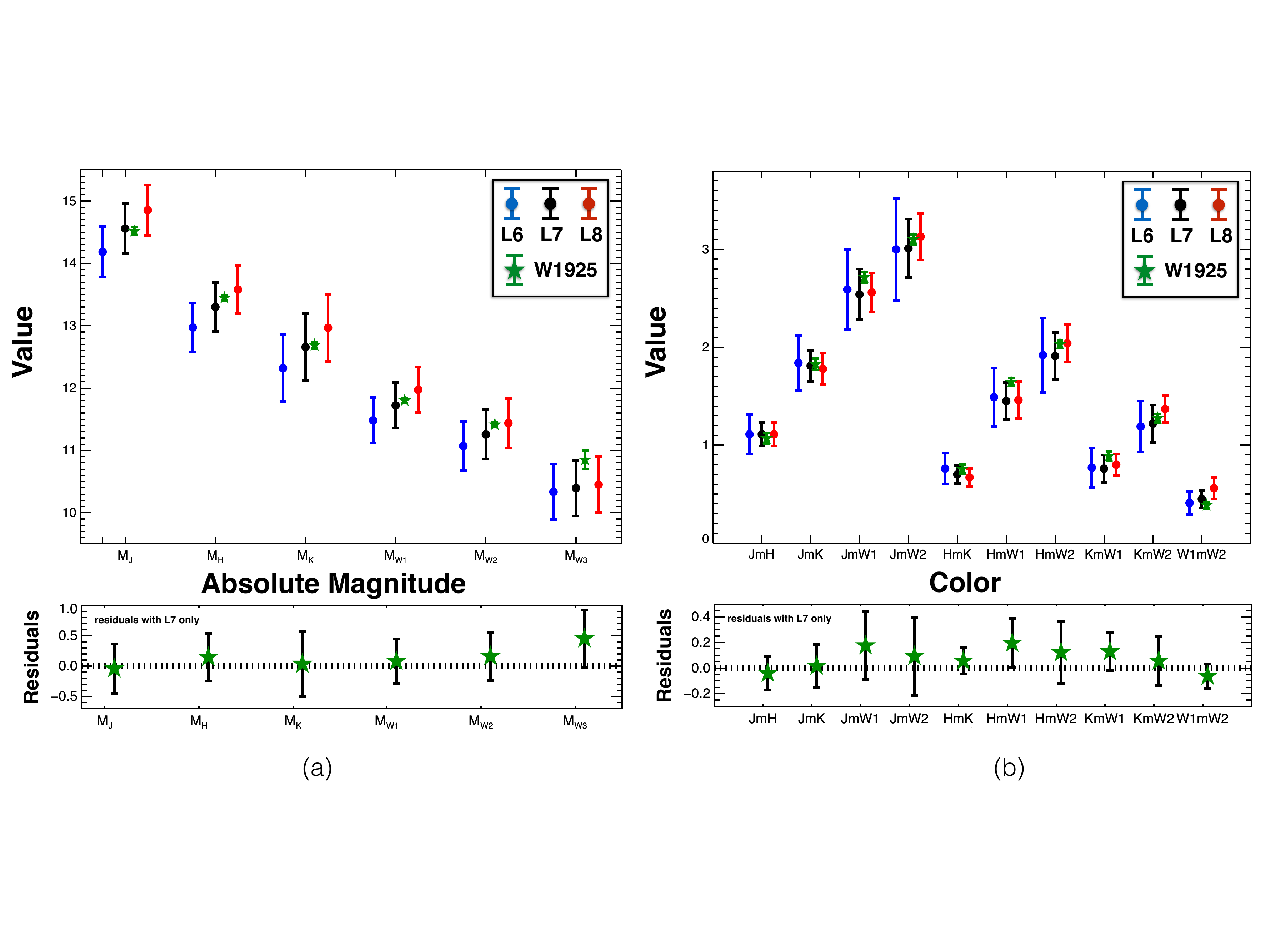}
\end{center}
\caption{{\bf Panel (a)} Absolute magnitudes of L6 (blue), L7 (black), and L8 (red) brown dwarfs across 2MASS and \emph{WISE} photometry from the polynomial relations reported in \citet{Faherty16}.  We plot the absolute magnitudes of W1925 and their uncertainties in green.  Residuals from just the L7 brown dwarfs are plotted in the bottom panel.  W1925 fits within the normal sources across all magnitudes.
{\bf Panel (b)} Average colors and spreads of L6 (blue), L7 (black), and L8 (red) dwarfs across 2MASS and \emph{WISE} from \citet{Faherty16}.  The colors of W1925 and their uncertainties are shown as green stars.  Residuals from just the L7 brown dwarfs are plotted along the bottom panel. W1925 fits well within the distribution of L7 dwarfs, although it is slightly red in $J-W1$, $H-W1$ and $K_s-W1$ colors.\label{fig:magscolors}} 
\end{figure*}

Figure~\ref{fig:magscolors} shows the average and spread of near- and mid-infrared colors for L6, L7, and L8 brown dwarfs from \citet{Faherty16}, compared to the measurements for W1925. All 2MASS and \emph{WISE} colors are consistent with the L7 spectral type. The bottom panel of Fig.~\ref{fig:magscolors} shows the residuals of each color against the distribution for L7 dwarfs.  The $(J-W1)$, $(H-W1)$, and $(K_s-W1)$ colors for W1925 are all on the red end of these distributions, which may be due to the atmosphere being slightly more cloudy than other L7 dwarfs or due to contamination from a nearby star. 

Using the \emph{Gaia}~DR2 parallax, we computed the absolute magnitudes of W1925 in 2MASS $J$, $H$, and $K_s$ bands, and \emph{WISE} $W1$, $W2$ and $W3$ bands. Figure~\ref{fig:magscolors} compares these absolute magnitudes to those of L6, L7, and L8 dwarfs calculated using the polynomial relations in~\citet{Faherty16}. Again, we find that W1925 aligns well with the late L dwarf field sequence, with no indication of unresolved binarity or other brightness anomalies.

\subsection{Kinematics}
\emph{Gaia}~DR2 reports a total proper motion of $\mu$\,=\,219.8\,$\pm$\,1.8\,mas\,yr$^{-1}$ and a parallax of $\varpi$\,=\,89.3\,$\pm$\,0.7\,mas for W1925 which yields a tangential velocity of $v_{\rm tan}$\,=\,11.67\,$\pm$\,0.13\,km\,s$^{-1}$.  The average $v_{\rm tan}$ value for L7 objects in \citet{Faherty09} was 30\,$\pm$\,9\,km\,s$^{-1}$,
placing W1925 on the low velocity end of this distribution. Including the measured radial velocity of -9\,$\pm$\,7\,km\,s$^{-1}$ 
implies UVW velocities of (-14.1\,$\pm$\,5.1, +2.2\,$\pm$\,4.8, 3.93\,$\pm$\,0.53)\,km\,s$^{-1}$ consistent with a field disk star. As a further check, we ran the full kinematics through the BANYAN~$\Sigma$ tool (\citealt{Gagne18}) and found 0\% probability for membership in any known moving group. 
Using the velocity distributions and population normalizations from \citet{Bensby03}, we find that this velocity vector corresponds to population probabilities of 93.9\% (thin disk), 6.1\% (thick disk), and 0.027\% (halo). We conclude that W1925 is most likely a brown dwarf member of the thin disk. 

\subsection{Fundamental Parameters\label{sec:fundamentals}}
Using the technique described in \citet{Filippazzo15}, we used the \emph{Gaia} parallax and 2MASS, \emph{WISE}, and Pan-STARRS photometry to construct a distance-calibrated spectral energy distribution for W1925 and compute a bolometric luminosity (log[$L_{\rm bol}$/$L_{\odot}$] = -4.443\,$\pm$\,0.008).  Using model-predicted radii measurements for 500 Myr - 10 Gyr objects, we semi-empirically calculated the effective temperature ($T_{\rm eff}$), mass, and $\log{g}$.  We list all calculated parameters in Table~\ref{tab:param}. Using the $L_{\rm bol}$ polynomial relations for field objects in \citet{Faherty16}, an L7 should have log($L_{\rm bol}$/$L_{\odot}$)\,=\,-4.426\,$\pm$\,0.133\,dex, hence W1925 is well within normal.  Similarly, the $T_{\rm eff}$ for W1925 fits well within the predicted field L7 value of 1401$\pm$113\,K.   At an estimated mass of 53$\pm$19\,$M_{\rm Jup}$, W1925 is between the deuterium and hydrogen burning minimum masses and should be classified as a brown dwarf.  
\section{Discussion}
Using the current sample of spectroscopically classified L0--L7 dwarfs\footnote{See https://jgagneastro.wordpress.com/list-of-ultracool-dwarfs/}, we estimate there are 40 sources known within $\sim$\,12\,pc of the Sun; 28 had trigonometric parallaxes before \emph{Gaia}~DR2, the remaining 12 have distances estimated from photometry and spectral type.  W1925 is the 30th closest L dwarf discovered to date. 

We can estimate the number of L dwarfs hiding in the Galactic plane by calculating a space density of known sources and applying it to the volume occupied by the plane. Looking at the collection of known objects, the majority of proper motion and photometric surveys have avoided $\sim$\,$|b|$\,$<$\,15\,degrees when targeting candidates.  Thus by (1) counting the number of spectroscopically confirmed L dwarfs outside the Galactic plane (109), (2) calculating the fraction of the total 12\,pc volume that is within the Galactic plane ($\sim$ 26\%), (3) constructing a probability density function of the space density to predict the number inside the Galactic plane with Bayes’ theorem based on a Poisson likelihood and a non-informative prior on the space density, and (4) subtracting the number of expected brown dwarfs in the Galactic plane to the number that is actually detected (22; and including Poisson error bars on this number), yields a predicted 3--12 L dwarfs within 12\,pc that remain to be detected in the Galactic plane at a 68\% confidence level. While this number is small, \emph{Gaia}~DR2 is an excellent resource for filling in the missing objects.

W1925 is an L/T transition dwarf and such sources have been found to show large amplitude photometric variability in the optical and infrared (e.g. \citealt{Gillon13},  \citealt{Radigan12}, \citealt{Metchev15}).  Given that W1925 is located in a dense, crowded area of the sky, it has numerous calibrator stars that can allow the source to be monitored with exceptionally high precision, making it an excellent target for future variability studies. The density of the field around W1925 also makes it an excellent candidate for adaptive optics imaging to search for faint companions that contributed negligibly to the overall luminosity of the system. Both of these benefits also apply to future brown dwarf discoveries in the Galactic plane.

\section{Conclusions}
We have spectroscopically confirmed W1925 to be an L7 brown dwarf at $11.20_{-0.08}^{+0.09}$\,pc.  We estimate a radial velocity of -9\,$\pm$\,7\,km\,s$^{-1}$ and calculate $UVW$ space velocities of (-14.1\,$\pm$\,5.1, +2.2\,$\pm$\,4.8, 3.93\,$\pm$\,0.53)\,km\,s$^{-1}$ respectively.  Using its spectrum, colors and absolute magnitudes, we confirm that W1925 has field brown dwarf properties, and likely has an age typical of field stars. From the parallax, photometry, and spectral data we create a spectral energy distribution and empirically determine log($L_{\rm bol}$/$L_{\odot}$)\,=\,-4.426\,$\pm$\,0.133\,dex. We translate these into semi-empirical estimates of $T_{\rm eff}$\,=\,1404\,$\pm$\,71\,K, $\log{g}$\,=\,5.1\,$\pm$\,0.4\,dex, and a mass of 53\,$\pm$\,19\,$M_{\rm Jup}$.  All values are consistent with equivalent spectral type (L7) objects in the field.

The discovery of W1925 is a demonstration of the power of \emph{Gaia} astrometry for mining the solar neighborhood, even in the most crowded regions of the sky near the Galactic plane.  We estimate there are a roughly 3--12 L-type brown dwarfs remaining to be found in this region, which constitutes 10--30\% of the overall population.  Finding such objects will not only complete the nearby stellar census, but will also serve as key targets for follow-up variability and multiplicity studies given the dense field of calibration stars.

\clearpage
\startlongtable
\begin{deluxetable}{cccl}
\tablecaption{Parameters for WISE~J192512.78+070038.8\label{tab:param}}
\tablehead{
\colhead{Parameter} & \colhead{Value} & \colhead{Units} & \colhead{Reference}}
\colnumbers
\startdata
{\bf \emph{Gaia}~DR2 ASTROMETRY}\\
\hline
\hline
$\alpha$ & 291.30335823223\tablenotemark{a} ($\pm$0.7\,mas) & deg & 1 \\
$\delta$ & +07.01101136280\tablenotemark{a} ($\pm$0.6\,mas) & deg & 1 \\
$\ell$\tablenotemark{a}    & 43.0438	& deg & 1 \\
$b$\tablenotemark{a}       & -4.2467  & deg & 1\\
$\varpi$       & 89.3\,$\pm$\,0.7 & mas & 1\\
$\mu_{\alpha}$ & 44.9\,$\pm$\,1.4  & mas\,yr$^{-1}$ & 1\\
$\mu_{\delta}$ & 215.2\,$\pm$\,1.2  & mas\,yr$^{-1}$ & 1\\
\hline
\hline
{\bf \emph{Gaia}~DR2 PHOTOMETRY}\\
\hline
\hline
$G_{BP}$ & 21.0$\pm$0.6 & mag &  1\\
$G$     & 20.038$\pm$0.009 & mag &  1\\
$G_{RP}$ & 18.20$\pm$0.04 & mag &  1\\
\hline
\hline
{\bf RELEVANT \emph{Gaia}~DR2 FLAGS}\\
\hline
\hline
Astrom. Excess Noise& 2.11 & mas & 1\\
Vis. Periods& 13 & $\cdots$ & 1\\
$\sigma$ 5D&  1.49 & mas & 1\\
\hline
\hline
{\bf OTHER PHOTOMETRY}\\
\hline
\hline
$r$ & 22.03$\pm$0.15 & mag & 2\\
$i$ &  20.149$\pm$0.015 &  mag &2\\
$z$ &  17.859$\pm$0.012 &  mag &2\\
$y$ &  16.868$\pm$0.010 &  mag &2\\
$J$ & 14.76$\pm$0.05  &  mag &3\\
$H$ & 13.69$\pm$0.03  &  mag &3\\
$K_s$ & 12.94$\pm$0.04  &  mag &3\\
$W1$\tablenotemark{b} & 12.005$\pm$0.023  &  mag &4\\
$W2$\tablenotemark{b} & 11.638$\pm$0.021  &  mag &4\\
$W3$\tablenotemark{b} & 10.93$\pm$0.11  &  mag &4\\
$W4$\tablenotemark{b} & $<$9.102   &  mag &4\\
\hline
\hline
{\bf SPECTROSCOPY}\\
\hline
\hline
Spectral Type (IR) & L7$\pm$1 & $\cdots$ &5\\
RV & $-$9$\pm$7 & km\,s$^{-1}$ & 5 \\
\hline
\hline
{\bf FUNDAMENTAL PARAMETERS}\\
\hline
\hline
log($L_{\rm bol}$/$L_{\odot}$) & -4.443$\pm$0.008 & $\cdots$ &5\\
$T_{\rm eff}$ & 1404$\pm$71 & K &5\\
Radius        & 0.99$\pm$0.10 & $R_{\rm Jup}$ & 5\\
Mass          & 53$\pm$19  & $M_{\rm Jup}$&5\\
$\log{g}$        & 5.1$\pm$0.4 & $\cdots$ & 5\\
\hline
\hline
{\bf CALCULATED KINEMATICS}\\
\hline
\hline
Distance\tablenotemark{c}       & 11.20$^{+0.09}_{-0.08}$ & pc & 5\\
$v_{\rm tan}$\tablenotemark{d} & 11.67\,$\pm$\,0.13 & km\,s$^{-1}$ & 5\\
$X$ & 8.16$\pm$0.06 &pc & 5\\
$Y$ & 7.62$\pm$0.06 &pc & 5\\
$Z$ & -0.83$\pm$0.01 &pc & 5\\
$U$ & -14.1$\pm$5.1  & km\,s$^{-1}$ & 5\\
$V$ & +2.2$\pm$4.8  &km\,s$^{-1}$ & 5\\
$W$ & +3.93$\pm$0.53 &km\,s$^{-1}$ & 5\\
\hline
\hline
{\bf ABSOLUTE MAGNITUDES}\\
\hline
\hline
M$_{J}$  & 14.52$\pm$0.06& mag & 5\\
M$_{H}$  & 13.45$\pm$0.05& mag & 5\\
M$_{K}$  & 12.69$\pm$0.05& mag & 5\\
M$_{W1}$ & 11.80$\pm$0.04& mag & 5\\
M$_{W2}$ & 11.41$\pm$0.04& mag & 5\\
M$_{W3}$ & 10.85$\pm$0.15& mag & 5\\
M$_{G}$    & 19.79$\pm$0.02& mag & 5\\
\enddata
\tablenotetext{a}{epoch J2015.5, ICRS}
\tablenotetext{b}{We chose the original \emph{WISE} catalog values in the analysis over the AllWISE values so we could compare to the photometry in \citet{Faherty16}}
\tablenotetext{c}{Calculated using D = 1/$\varpi$, which is good approximation for parallax known to $\varpi$/$\sigma_{\varpi}$ = 133 accuracy}
\tablenotetext{d}{Calculated using \citet{GaiaDR2} astrometry.}
\tablecomments{The object does not have entries in GSC 2.2, USNO-B1.0, and does not appear on any of the photographic sky surveys scanned by SuperCOSMOS. \\
References: (1) \citet{GaiaDR2}, (2) \citet{Chambers16}, (3) \citet{Cutri03}, (4) \citet{Wright10}, (5) This paper.}
\end{deluxetable}

\acknowledgments

This work has made use of data from the European Space Agency (ESA)
mission {\it \emph{Gaia}} (\url{https://www.cosmos.esa.int/gaia}), processed by
the {\it \emph{Gaia}} Data Processing and Analysis Consortium (DPAC,
\url{https://www.cosmos.esa.int/web/gaia/dpac/consortium}). Funding
for the DPAC has been provided by national institutions, in particular
the institutions participating in the {\it \emph{Gaia}} Multilateral Agreement.
AJB acknowledges funding support from the
National Science Foundation under award No.\ AST-1517177
ECG acknowledges funding support from the
National Science Foundation under award No.\ AST-1313278
EEM acknowledges support from the NASA NExSS program and a JPL RT\&D award.
FM was supported by an appointment to the NASA Postdoctoral Program at the Jet Propulsion Laboratory, administered by Universities Space Research Association under contract with NASA.
We thank the staff of Palomar Observatory, including Kevin Rykoski, Kajsa Peffer, and Paul Nied, for a successful night of observing with TripleSpec.
JF thanks Ricky Smart for useful conversations about Galactic plane contamination and the Fellmans for writing accommodations. 
Part of this research was carried out at the Jet Propulsion Laboratory, California Institute of Technology, under a contract with the National Aeronautics and Space Administration.
This research was started at the NYC Gaia DR2 Workshop at the Center for Computational Astrophysics of the Flatiron Institute in 2018 April.

\vspace{5mm}
\facilities{\emph{Gaia}, Hale(TripleSpec), \emph{WISE}, CTIO:2MASS, \emph{UKIRT}}

\software{Aladin, BANYAN~$\Sigma$ (\citealt{Gagne18})}

\end{document}